\documentclass[useAMS,usenatbib,a4paper,referee]{mn2e}
\usepackage{graphicx}
\usepackage{amsmath}
\usepackage{txfonts}
\usepackage{mathrsfs}

\def\aap{A\&A}
\def\apjl{ApJ}

\def\apjs{ApJS}

\newcommand{\be}{\begin{equation}}
\newcommand{\ee}{\end{equation}}
\newcommand{\bea}{\begin{eqnarray}}
\newcommand{\eea}{\end{eqnarray}}

\title[Cosmic Chronometers]{Cosmic Chronometers in the $R_h=ct$ Universe}
\author[Fulvio Melia and Robert S. Maier]{Fulvio Melia$^{1}$\thanks{John Woodruff Simpson 
Fellow. E-mail: fmelia@email.arizona.edu} and Robert S. Maier$^{2}$\thanks{E-mail: rsm@math.arizona.edu}\\
\null$^{1}$Department of Physics, The Applied Math Program, and Department of Astronomy, 
The University of Arizona, AZ 85721, USA\\
\null$^{2}$Department of Mathematics, The Statistics Program, and Department of Physics,
The University of Arizona, AZ 85721, USA}

\begin{document}

\date{}

\pagerange{\pageref{firstpage}--\pageref{lastpage}} \pubyear{2013}

\maketitle

\label{firstpage}

\begin{abstract}
The use of luminous red galaxies as cosmic chronometers provides us with an
indispensable method of measuring the universal expansion rate $H(z)$ in a
model-independent way.  Unlike many probes of the cosmological history,
this approach does not rely on integrated quantities, such as the
luminosity distance, and therefore does not require the pre-assumption of
any particular model, which may bias subsequent interpretations of the
data.  We employ three statistical tools -- the Akaike, Kullback, and Bayes
Information Criteria (AIC, KIC and BIC) -- to compare the $\Lambda$CDM model
and the $R_{\rm h}=ct$ Universe with the currently available measurements
of $H(z)$, and show that the $R_{\rm h}=ct$ Universe is favored by these
model selection criteria.  The parameters in each model are individually
optimized by maximum likelihood estimation.  The $R_{\rm h}=ct$ Universe
fits the data with a reduced $\chi^2_{\rm dof}=0.745$ for a Hubble constant
$H_0=63.2\pm2.5$ km s$^{-1}$ Mpc$^{-1}$, and $H_0$~is the sole parameter in
this model.  By comparison, the optimal $\Lambda$CDM model, which has three
free parameters (including $H_0=68.9\pm 2.4$ km s$^{-1}$ Mpc$^{-1}$,
$\Omega_m=0.32$, and a dark-energy equation of state $p_{\rm de}=-\rho_{\rm
  de}$), fits the $H(z)$ data with a reduced $\chi^2_{\rm dof}=0.777$.
With these $\chi^2_{\rm dof}$ values, the AIC yields a likelihood of
$\approx 82$ per cent that the distance--redshift relation of the $R_{\rm h}=ct$
Universe is closer to the correct cosmology, than is the case for
$\Lambda$CDM\null.  If the alternative BIC criterion is used, the
respective Bayesian posterior probabilities are 91.2 per cent ($R_{\rm h}=ct$)
versus $8.8$ per cent ($\Lambda$CDM).  Using the concordance $\Lambda$CDM
parameter values, rather than those obtained by fitting $\Lambda$CDM to the
cosmic chronometer data, would further disfavor $\Lambda$CDM.
\end{abstract}


\section{Introduction}
The expansion of the Universe is now being studied by several methods,
including observations of Type~Ia SNe (Riess et~al.\ 1998; Perlmutter
et~al.\ 1999), weak lensing (Refregier 2003), baryon acoustic oscillations
(Seo \& Eisenstein 2003; Eisenstein et~al.\ 2005; Pritchard
et~al.\ 2007; Percival et~al.\ 2007), and cluster counts (Haiman
et~al.\ 2000), among several others.  Each of these methods presents its
own set of difficulties, among them a dependence on integrated quantities,
such as the luminosity distance which, however, is not independent of the
assumed cosmology.  It is therefore quite difficult to use the data for
unbiased, comparative studies to test different expansion histories.  This
problem is particularly acute in the case of Type~Ia SNe, where at least
four `nuisance' parameters characterizing the standard candle must be
optimized simultaneously with the model's free parameters, rendering the
data compliant to the underlying cosmology (see, e.g., Melia 2012a).

Even so, some progress has been made recently with attempts at comparing
predictions of the $R_{\rm h}=ct$ Universe (Melia 2007; Melia \& Shevchuk
2012) with the data, and with $\Lambda$CDM\null.  The evidence thus far
seems to suggest that the $R_{\rm h}=ct$ cosmology is a better match to the
observations at high redshifts, particularly when it comes to the
large-scale fluctuations of the cosmic microwave background (CMB),
expressed through its angular correlation function and the apparent
alignment of its quadrupole and octopole moments (for a summary of these
comparisons, see Melia 2012a).

In the local universe, these two models are virtually indistinguishable,
e.g., in predicting a very similar luminosity distance all the way out to a
redshift of~6 and beyond.  Thus, given the problem of identifying
model-independent data through Type~Ia supernova observations, it is not
easy to evaluate one model against the other on the basis of these
measurements alone.  However, some clarification begins to emerge beyond a
redshift of~6, where the high-$z$ quasars are now known to be accreting at,
or near, their Eddington limit (see, e.g., Willott et~al.\ 2010a,b; De~Rosa
et~al.\ 2011).  We showed recently that a Hubble Diagram (HD) constructed
from these sources reveals a cosmic expansion fully consistent with the
$R_{\rm h}=ct$ Universe, assuming a current value of $69\pm 4$ km s$^{-1}$
Mpc$^{-1}$ for the Hubble constant $H_0$ (Melia 2012b).  Interestingly,
$\Lambda$CDM can also fit the high-$z$ quasar HD, but only for a very
specific set of parameters, including a matter energy density
$\Omega_m=0.27$, scaled to its current critical value.  But whereas the
$R_{\rm h}=ct$ Universe has only one free parameter -- the Hubble
constant -- the standard model has as many as six (depending on how one
parametrizes the dark-energy equation of state $w_{\rm de}\equiv p_{\rm
  de}/\rho_{\rm de}$) -- including $H_0$, $\Omega_m$, and $w_{\rm de}$.  The
implication of this is that the optimization of $\Lambda$CDM simply forces
it to relax to the $R_{\rm h}=ct$ expansion profile, which is more robust.

Moreover, though the distance--redshift relationship is essentially the
same in these two models (even out to $z\ga 6-7$), the age--redshift
relationship is not.  In fact, these same high-$z$ quasars present a
seemingly insurmountable problem for $\Lambda$CDM because they suggest that
$\sim 10^9\;M_\odot$ supermassive black holes appeared only 700--900 Myr
after the big bang.  Instead, in $R_{\rm h}=ct$, their emergence at
redshift $\sim 6$ corresponds to a cosmic age of $\ga 1.6$ Gyr.  This was
enough time for them to begin growing from $\sim 5-20\; M_\odot$ seeds
(presumably the remnants of Pop~II and~III supernovae) at $z\la 15$ (i.e.,
\emph{after} the onset of re-ionization) and still reach a billion solar
masses by $z\sim 6$ via standard, Eddington-limited accretion (Melia 2013).

This kind of tangible result suggests that the $R_{\rm h}=ct$ Universe
relieves the growing tension between $\Lambda$CDM and the observations, but
it would be highly beneficial for us to find a way of testing this
cosmology -- and quantifying its superiority over $\Lambda$CDM -- by
exploiting model-independent data in the nearby Universe.  The purpose of
this paper is to demonstrate that the use of luminous red galaxies as
cosmic chronometers (Jimenez \& Loeb 2002) can do just that.  We shall show
that over the redshift range $0\la z\la 1.8$, the measured Hubble constant
$H(z)$ is fitted better by the $R_{\rm h}=ct$ model than by $\Lambda$CDM;
and especially so, if one takes account of the reduction in the number of
free parameters.  Unlike other indicators that rely on the expansion
history of the Universe, the cosmic chronometers may therefore offer us the
best evidence yet that $R_{\rm h}=ct$ is to be preferred over $\Lambda$CDM.

We introduce the cosmic chronometers in~\S\,2, and in~\S\,3 discuss the
AIC, KIC and BIC tools we use to test $\Lambda$CDM and the $R_{\rm h}=ct$
Universe against these data.  The results of our comparison between
$\Lambda$CDM and $R_{\rm h}=ct$ are presented in~\S\,4 and discussed
in~\S~5.  We conclude in~\S~6 with a discussion of future prospects.

\section{The Cosmic Chronometers}
Cosmic chronometers offer us the possibility of measuring the differential
age of the Universe, circumventing the limitations associated with the use
of integrated histories, by directly measuring the derivative ${\rm d}t/{\rm d}z$,
which represents the change in cosmic time as a function of redshift.  And
since $H(z)\equiv \dot{a}/a$, in terms of the expansion factor $a(t)$, a
measurement of ${\rm d}t/{\rm d}z$ directly yields the expansion rate, because
\begin{equation}
H(z)={\dot{a}\over a}=-\,{1\over 1+z}{{\rm d}z\over {\rm d}t}\;.
\end{equation}

For various reasons, the best cosmic chronometers appear to be galaxies
that are evolving passively on a time-scale much longer than their age
difference.  Observations indicate that the most massive galaxies contain
the oldest stellar populations up~to redshifts $z\sim 1-2$ (Dunlop
et~al.\ 1996; Spinrad et~al.\ 1997; Cowie et~al.\ 1999; Heavens
et~al.\ 2004; Thomas et~al.\ 2005).  Less than $1$ per cent of the stellar mass in
these massive galaxies formed at $z<1$ (Heavens et~al.\ 2004; Panter
et~al.\ 2007).  In high-density regions (i.e., galaxy clusters), star
formation ceased by redshift $z\sim 3$ (Thomas et~al.\ 2005), and other
massive systems -- those with stellar masses $\ga 5\times 10^{11}\;
M_\odot$ -- finished their star formation activity by $z\sim 2$ (Treu
et~al.\ 2005).

The empirical evidence therefore suggests that galaxies in the highest
density regions of clusters formed their stellar content at $z\ga 2$, and
have been evolving passively since that time, without any additional
episodes of star formation.  One can therefore view these galaxies as
tracing the `red envelope,' hosting the oldest stars in the Universe at
every redshift.  Thus, given their viability as cosmic chronometers, a
great deal of effort is being expended to calculate ${\rm d}t/{\rm d}z$ -- and therefore
$H(z)$ -- using their measured properties (see, e.g., Stern et~al.\ 2010;
Stern et~al.\ 2012; Moresco et~al.\ 2012a,b).

For example, one of the most direct ways of determining the age of the
galaxy is to use the 4000\,\AA\ break in its spectrum, which depends
linearly on age for old stellar populations (Moresco et~al.\ 2011).  This
break is a discontinuity of the spectral continuum due to metal absorption
lines whose amplitude correlates linearly with the age and metal abundance.
If the metallicity is known, then the difference in age between two
galaxies is proportional to the difference in their 4000\,\AA\ amplitudes.

However, one must also be aware of the fact that many systematic sources of
uncertainty can bias this kind of analysis (see, e.g., Moresco et~al.
2012b).  These include: (1)~the degeneracy between the effect of a change in
age and an effect due to a change in stellar metallicity or the star
formation history; (2)~the possible biasing of the estimate of $H(z)$ by
the choice of stellar population synthesis model, used to estimate the age
or calibrate the 4000\,\AA\ versus age relation; and (3)~the possible
existence of a progenitor bias (van~Dokkum \& Franx 1996), in which
high-redshift samples of early-type galaxies might not be statistically
equivalent to those at low redshifts.

These caveats notwithstanding, one is none the less encouraged by the
agreement seen between the results of several different approaches.  The
data set shown in figure~1, including both $H(z)$ measurements and error
bars, was assembled from the compilations of Simon et~al.\ (2005), Stern
et~al.\ (2010), and Moresco et~al.\ (2012a), and spans the redshift range
$0\le z\la 1.8$.  Together, these compilations paint a fairly consistent
picture of the universal expansion, particularly when viewed in terms of
the theoretical expectations, which we shall consider shortly, following
our discussion of model selection statistics.

\section{Model Selection Statistics}
To compare the evidence for and against competing models, such as models of
the distance--redshift relationship, the use of the Akaike Information
Criterion (AIC) is now common in cosmology (see, e.g., Takeuchi 2000;
Liddle 2004, 2007; Tan \& Biswas 2012).  The AIC can be viewed as an
enhanced `goodness of fit' criterion, which extends the usual~$\chi^2$
criterion by taking account of the number of parameters in each model.  It
prefers models with few parameters to those with many, unless the latter
provide a substantially better fit to the data.  This reduces the
possibility of overfitting: the fact that by optimizing over a greater
number of parameters, one may simply be fitting the noise.

As developed (Akaike 1973; see also Burnham \& Anderson 2002, 2004), the
AIC provides the relative ranks of two or more competing models, and also a
numerical measure of confidence that each model is the best.  These
confidences are analogous to likelihoods or posterior probabilities in
traditional statistical inference.  But unlike traditional inference
methods, the AIC can be applied to models that are not `nested.'
Comparing a pair of models that are nested, in the sense that one is a
specialization of the other, is straightforward: after fitting each model
to the data, one computes the $\chi^2$ per degree of freedom for each, and
decides which is a better fit.  One can also calculate (say, by applying an
F-test) a likelihood that the simpler model should be rejected, or the
likelihood of the null hypothesis that the simpler model is a better
approximation to the `true' one.  By exploiting the AIC one can
generalize this procedure: one can compare a pair of models, neither of
which is a specialization of the other; such as $\Lambda$CDM and an
alternative model.

The AIC can be applied after regression of the following kind is performed.
Suppose that for values $z_1,\dots,z_n$ of an independent variable there
are measured values $h_1,\dots,h_n$ of a dependent one, with (known) error
bars $\pm\sigma_1,\dots,\pm\sigma_n$; and suppose the errors are normally
distributed.  Suppose that a model~$\mathcal{M}$ predicts values $\hat
h_1,\dots,\hat h_n$, computed from a formula $\hat h_i=\hat h_i(\vec\beta)$
that involves a parameter vector~$\vec{\beta}$ comprising $k$~unknown
parameters, i.e., $\vec{\beta}=(\beta_1,\dots,\beta_k)$.  That is, the data
model~$\mathcal{M}$ is really a statistical one, of the form
\begin{equation}
\label{eq:firstAIC}
  h_i = \hat h_i({\vec\beta}) + \sigma_i Z_i\;,
\end{equation}
where $Z_1,\dots,Z_n$ are independent standard normal random variables.
(In the case of linear regression, $\hat h_i(\vec\beta)$~would be
$\sum_{j=1}^k X_{ij}\beta_j$ for known coefficients~$X_{ij}$; typically,
$X_{ij}=\hat h^{(j)}(z_i)$ for known functions $\hat h^{(1)},\dots,\hat
h^{(k)}$ of~$z$.)

For model~$\mathcal{M}$, the $\chi^2$ goodness of its fit to the data is
given by
\begin{equation}
\label{eq:secondAIC}
\chi^2 = \sum_{i=1}^n [h_i - \hat h_i(\vec\beta)]^2/\sigma_i^2\;,  
\end{equation}
i.e., a (weighted) sum of squared errors, and the reduced $\chi^2$ (i.e.,
the $\chi^2$ per degree of freedom) by
\begin{equation}
\chi^2_{\rm{dof}} = \chi^2 / (n-k)\;.
\end{equation}
(It is assumed that $n>k$.)  The parameters $(\beta_1,\dots,\beta_k)$ are
chosen to minimize the~$\chi^2$, yielding the best fit to the data.  The
AIC for the resulting fitted model is then given by
\begin{equation}
\label{eq:AIC}
  {\rm AIC} = \chi^2 + 2\,k\;.
\end{equation}
If there are two or more competing models for the data,
$\mathcal{M}_1,\dots,\mathcal{M}_N$, and they have been separately fitted,
the one with the least resulting AIC is assessed as the one most likely to
be nearest to the `truth,' i.e., to the unknown model~${\cal M}_*$ that
generated the data.  A more quantitative ranking of models can be computed
as follows.  If ${\rm AIC}_\alpha$ comes from model~$\mathcal{M}_\alpha$,
the unnormalized likelihood that $\mathcal{M}_\alpha$~is closest to the
truth is the `Akaike weight' $\exp(-{\rm AIC}_\alpha/2)$.  Informally,
$\mathcal{M}_\alpha$~has likelihood
\begin{equation}
\label{eq:lastAIC}
{\cal L}(\mathcal{M}_\alpha)=
\frac{\exp(-{\rm AIC}_\alpha/2)}
{\exp(-{\rm AIC}_1/2)+\dots+\exp(-{\rm AIC}_N/2)}
\end{equation}
of being the best choice.  (The $2$'s here could of~course be omitted by
redefining ${\rm AIC}$, but the normalization implicit in Equation~(5) is
traditional.)  In the case of a pair of models
$\mathcal{M}_1,\mathcal{M}_2$, the difference ${\rm AIC}_2\nobreak-{\rm
  AIC}_1$ determines the extent to which $\mathcal{M}_1$ is favored
over~$\mathcal{M}_2$.

It is clear that the $2k$ term in Equation~(5), proportional to the
parameter count~$k$, exponentially disfavors models with too many
parameters, though such models can be favored if they do a much better job
of fitting the data.  The choice of proportionality constant (i.e.,~$2$) is
not entirely arbitrary, being based on an argument from information theory
that has close ties to statistical mechanics.  The following is a brief
summary, with many more details to be found in the statistics literature.
(The reader should note that most of the literature focuses on the case
when the error variances $\sigma_1^2,\dots,\sigma_n^2$ are both unknown and
equal to some common variance~$\sigma^2$, a nuisance parameter that must be
estimated as part of the fitting process; the setup given in Equations 2--3
is actually simpler.)

Any two statistical models of the data set $(h_1,\dots,h_n)$, such as a
`true' model $\mathcal{M}_*$ and another model~$\mathcal{M}$, can be
viewed as probability density functions (PDF's) on~$\mathbb{R}^n$, say
$f_*(h_1,\dots,h_n)$ and $f(h_1,\dots,h_n)$, respectively.  In information
theory one says that the discrepancy of the PDF $f$ from the PDF~$f_*$,
which is a measure of distance, is given by the Kullback--Leibler formula
\begin{equation}
  D(\mathcal{M}_*\|\mathcal{M}) = \int_{\mathbb{R}^n} 
{\rm d}h_1\dots {\rm d}h_n\, f_*(h)\,\ln\frac{f_*(h)}{f(h)}\,\ge\,0
\end{equation}
(where in an obvious notation, the argument $h$ stands for the entire
data set $[h_1,\dots,h_n]$).  In a thermodynamic interpretation this is a
relative entropy in the sense of Boltzmann and Hasen\"ohrl.  To select the
best model~$\mathcal{M}$ from a set of candidate models, one would choose
the one with the minimum $D(\mathcal{M}_*\|\mathcal{M})$.  Of~course
$\mathcal{M}_*$~is not known, so this cannot be done literally.  But the
case when $\mathcal{M}$~is a parametrized model, and its parameters are
chosen (by minimizing~$\chi^2$) to fit a data set generated
by~$\mathcal{M}_*$, is special.  It can be shown that the AIC of the fitted
model~$\mathcal{M}$ is a good approximation to
$2D(\mathcal{M}_*\|\mathcal{M})$, up~to an unimportant additive constant.
This is especially the case when $\mathcal{M}_*$ is a model of the same
type, with unknown parameters $(\beta_1^*,\dots,\beta_k^*)$.

Specifically, ${\rm AIC}/2$ is an unbiased estimator of the distance
$D(\mathcal{M}_*\|\mathcal{M})$: exactly so for linear regression, and to
leading order for non-linear regression.  The phrase `unbiased estimator'
means that \emph{on average} the two are the same, where the averaging is
over data sets generated by~$\mathcal{M}_*$, with PDF~$f_*$.  Of~course the
fitted model~$\mathcal{M}$ depends on the data set, so in the context
of~$\mathcal{M}_*$, both $D(\mathcal{M}_*\|\mathcal{M})$ and
$\textrm{AIC}/2$ are random variables.  In probabilistic language, the lack
of bias means that they have the same expectation.

The extent to which the fitted AIC is an \emph{accurate} estimate of
$2D(\mathcal{M}_*\|\mathcal{M})$, data set by data set, as well as being the
same on average, has been investigated theoretically (Yanagihara \& Ohmoto
2005).  Its variability has also been studied empirically; for example, by
repeatedly comparing $\Lambda$CDM to other cosmological models on the basis
of data sets generated by a bootstrap method (Tan \& Biswas 2012).  It is
known that the AIC is increasingly accurate when $n$~is large, but it is
felt that for all~$n$, the magnitude of the difference $\Delta=\allowbreak
{\rm AIC}_2 -\nobreak {\rm AIC}_1$ should provide a numerical assessment of
the evidence that model~1 is to be preferred over model~2.  A~rule of thumb
that has been used in the literature is that if $\Delta\la2$, the evidence
is weak; if $\Delta\approx3$ or~$4$, it is mildly strong; and if
$\Delta\ga5$, it is quite strong.

Besides using fixed thresholds, one can weight each candidate model in a
Boltzmann-like way by its Akaike weight, i.e., according to Equation~(6).
For each model $\mathcal{M}_\alpha$, the likelihood ${\cal
  L}(\mathcal{M}_\alpha)$, which is determined by the differences between
${\rm AIC}_\alpha$ and the AIC's of the other model(s), is loosely
analogous to a posterior probability in statistical inference, despite its
not being computed by a Bayesian procedure (no~Bayesian prior is involved).
But in the absence of a general theory of AIC variability, deciding between
models 1 and~2 cannot be viewed as a hypothesis test, at any fixed level of
significance such as~0.05.

Several alternatives to the AIC have been considered in the literature.  A
lesser-known one arises as follows.  The discrepancy
$D(\mathcal{M}_*\|\mathcal{M})$ is not symmetric in the PDF's $f_*,f$, and
it has been suggested that it should be replaced by a symmetrized version,
which is arguably a better tool for distinguishing between data models
(Cavanaugh 1999).  The unbiased estimator for the symmetrized version has
been given the name KIC (Kullback Information Criterion), and is given by
\begin{equation}
\label{eq:KIC}
  {\rm KIC} = \chi^2 + 3\,k\;.
\end{equation}
The KIC, with $k$ multiplied by the coefficient~3 rather than~2, disfavors
overfitting more than does the AIC, and has been shown to perform favorably
against the AIC as a tool for model selection (Cavanaugh 2004).  It has
long been felt (since, e.g., Bhansali \& Downham 1977) that the problem of
overfitting may be best dealt with by choosing a coefficient that is larger
than~2, and perhaps even than~3.  But the AIC and KIC are the only such
schemes that follow readily from information theory.

A better-known alternative to the AIC is the BIC (Bayes Information
Criterion), which is a misnomer in that it is not based on information
theory, but rather on an asymptotic ($n\to\infty$) approximation to the
outcome of a conventional Bayesian inference procedure for deciding between
models (Schwarz 1978).  It is defined by
\begin{equation}
\label{eq:BIC}
  {\rm BIC} = \chi^2 + (\ln n)\,k\;,
\end{equation}
and suppresses overfitting very strongly if $n$~is large.  Liddle
et~al.\ (2006) and Liddle (2007) make the case for using BIC in
cosmological model selection, and it has now been used to compare several
popular models against $\Lambda$CDM (see, e.g., Shi et~al.\ 2012).
However, it should be noted that the monograph of Burnham \& Anderson
(2002), which popularized Equation~(6) for assigning AIC-based likelihoods
to models, strongly prefers AIC to BIC as a tool for model selection\null.
They elsewhere note that the AIC can in~fact be interpreted in Bayesian
terms, as being the consequence of imposing a nonuniform but reasonable
choice of prior distribution on the set of candidate models (Burnham \&
Anderson 2004).  Kuha (2004) draws further analogies between AIC and BIC,
and argues that they are both valuable tools.

In the comparison below, we employ the AIC, KIC, and BIC\null.  We do not
employ the so-called corrected AIC, denoted ${\rm AIC}_{\rm c}$, which
includes a correction term intended to remove bias when $n$~is small
(Burnham \& Anderson 2002).  The correction term is small (cf.\ Tan \&
Biswas 2012).  More importantly, the form of this term is appropriate only
for data sets without explicit error bars, with the common error
variance~$\sigma^2$ estimated as part of the fitting process (Maier 2013,
in preparation).

\section{A Comparison between $\Lambda$CDM and $R_{\rm h}=ct$}
The $R_{\rm h}=ct$ Universe is a flat Friedmann--Robertson--Walker (FRW)
cosmology that strictly adheres to the constraints imposed by the
simultaneous application of the Cosmological principle and Weyl's postulate
(Melia, 2007; Melia \& Shevchuk 2012; Melia 2012a).  When these ingredients
are applied to the cosmological expansion, the gravitational horizon
$R_{\rm h}=c/H$ must always be equal to~$ct$.  This cosmology is therefore
very simple, because $a(t)\propto t$, which also means that $1+\nobreak
z=\allowbreak 1/t$, with the (standard) normalization that $a(t_0)=1$.
Therefore in the $R_{\rm h}=ct$ Universe, we have the straightforward
scaling
\begin{equation}
H(z)=(1+z)H_0\;.
\end{equation}
Notice, in particular, that the expansion rate $H(z)$ in this model has
only one free parameter.  By comparison, $\Lambda$CDM has as many as six
parameters (depending on the application), including~$H_0$, the scaled
matter energy density $\Omega_m$ ($\equiv \rho_{\rm m}/\rho_c$, in terms of
the matter energy density~$\rho_{\rm m}$ and the critical density
$\rho_c\equiv\allowbreak [3c^2/8\pi G]H_0^2$), and the dark-energy equation
of state $w_{\rm de}= p_{\rm de}/\rho_{\rm de}$.

In this paper, we shall take the minimalist approach and optimize
$\Lambda$CDM using only these three free parameters.  (Using additional
parameters would weaken the statistical significance of the fit even
further, so by selecting this minimal set, we present $\Lambda$CDM in its
best possible light.)  The Hubble constant in this cosmology is therefore
given by
\begin{equation}
H(z)=H_0\left[\Omega_{\rm m}(1+z)^3+\Omega_{\rm r}(1+z)^4+
\Omega_{\rm de}(1+z)^{3(1+w_{\rm de})}\right]^{1/2}\;,
\end{equation}
where $\Omega_{\rm r}$ and $\Omega_{\rm de}$ for radiation and dark energy,
respectively, are defined analogously to~$\Omega_{\rm m}$.  In addition, we
shall assume a flat $\Lambda$CDM cosmology, for which $\Omega_{\rm
  m}+\nobreak\Omega_{\rm r}+\nobreak\Omega_{\rm de}=1$, thus avoiding the
introduction of $\Omega_{\rm de}$ as an additional free parameter.
Of~course, $\Omega_{\rm r}$ ($\sim 6\times 10^{-5}$) is known from 
the current temperature ($\approx 2.7$ K$^\circ$) 
of the cosmic microwave background radiation.

For each model $\mathcal{M}_\alpha$ (with $\alpha=1,2$ specifying the
$R_{\rm h}=ct$ Universe and $\Lambda$CDM, respectively), we optimize the
fit by finding the model parameter vector~${\vec\beta}_\alpha$ that
minimizes the~$\chi^2$.  Equivalently, we choose~$\vec\beta_\alpha$ to
maximize the joint likelihood function
\begin{equation}
\Phi_\alpha(\vec{\beta}_\alpha)\equiv \prod_{i=1}^{n}{e^{-\left[H_i-H(z_i|\vec{\beta}_\alpha)\right]^2/2\sigma_i}
\over \sqrt{2\pi}\sigma_i}\;,
\end{equation}
where the $H_i$ are the measured values of the Hubble constant at
redshift~$z_i$, and the $H(z_i|\vec{\beta}_\alpha)$ are the corresponding
theoretical values computed from the parameter vector~$\vec{\beta}_\alpha$.
For $\alpha=1,2$, the number of parameters (i.e., the length~$k$ of the
vector~$\vec\beta_\alpha$) is respectively 1~and~3, as stated.  The fitting
is a linear regression in the case of the $R_{\rm h}=ct$ Universe and a
non-linear one for $\Lambda${CDM}, as is evident from Equations (10)
and~(11).  The data set $\{(z_i,H_i,\sigma_i)\}_{i=1}^n$ to which each model
is fitted was assembled from the $H(z)$ compilations of Moresco
et~al.\ (2012a), Stern et~al.\ (2010), and Simon et~al.\ (2005), and
consists of $n=19$ measured values, each with an error bar.

The results for the $R_{\rm h}=ct$ Universe (for which the best fit has
$H_0=63.2\pm 2.5$ km s$^{-1}$ Mpc$^{-1}$) and $\Lambda$CDM (for which it
has $H_0=68.9\pm2.4$ km s$^{-1}$ Mpc$^{-1}$, $\Omega_{\rm m}=0.32$, and
$w_{\rm de}=-1$) are shown in figure~1.  (These $H_0$ values are quoted
with one-sigma standard errors, calculated from the corresponding
$\chi^2$-distribution for each model.)  With $19\nobreak-1=\allowbreak18$
degrees of freedom, the reduced $\chi^2_{\rm dof}$ for the $R_{\rm h}=ct$
Universe is~0.745.  By comparison, the optimal $\Lambda$CDM fit has
$19\nobreak-3\allowbreak=16$ degrees of freedom, and a corresponding
reduced $\chi^2_{\rm dof}=0.777$.  Even by eye, one can see that $R_{\rm
  h}=ct$ is a better fit to the data at $z\ga 0.9$.  The reduced~$\chi^2$
overall suggests that $R_{\rm h}=ct$ is at least as good as $\Lambda$CDM;
especially, when its having only one parameter is taken into account.  We
shall see shortly that on statistical grounds, the $R_h=ct$
distance--redshift predictions are in~fact more likely than those of
$\Lambda$CDM to be closer to the correct cosmology.

\begin{figure}
\center{\includegraphics[scale=0.9,angle=0]{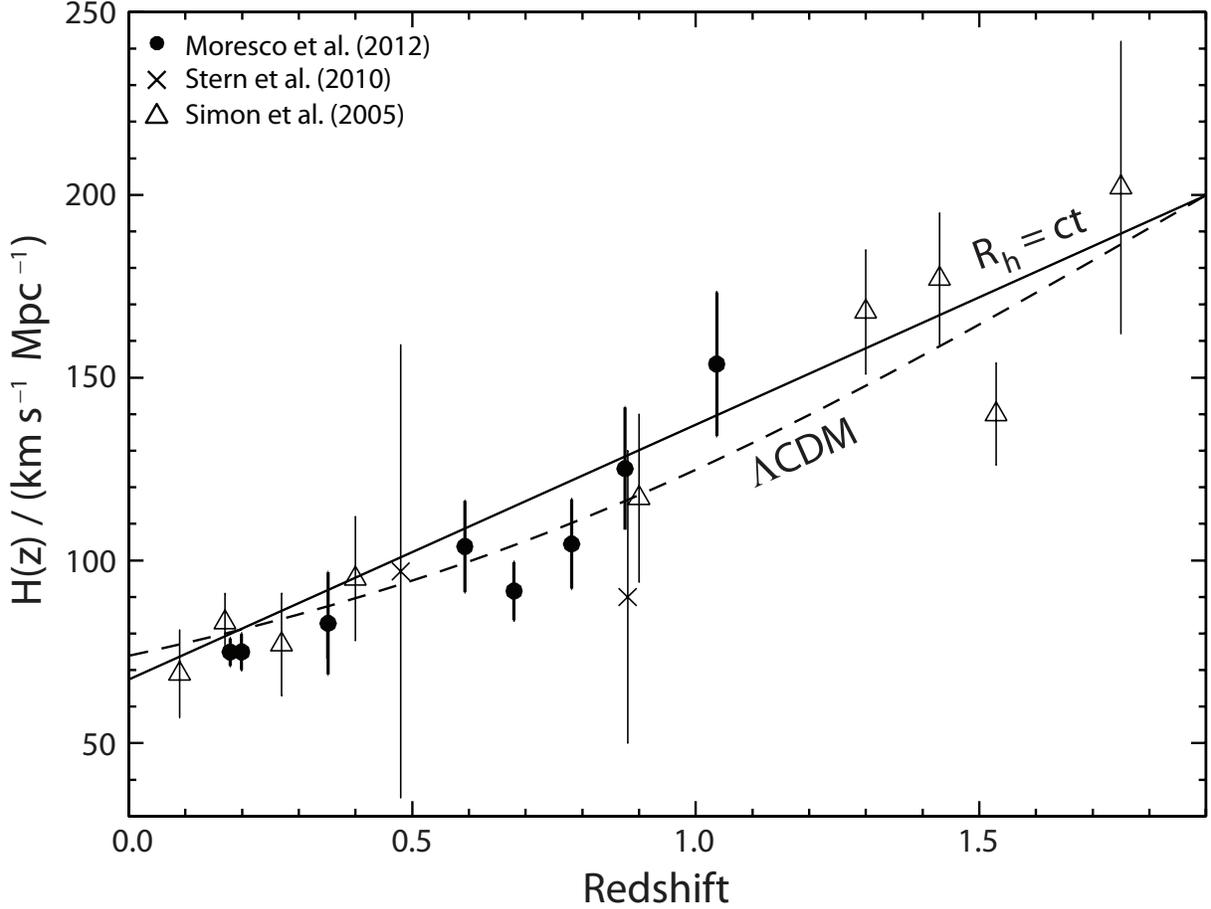}
\caption{Nineteen $H(z)$ measurements, with error bars, and comparison with two 
theoretical models: (\emph{solid}) the $R_{\rm h}=ct$ Universe, with its sole
parameter $H_0=63.2 \pm 2.5$ km s$^{-1}$ Mpc$^{-1}$, and (\emph{dashed})
the standard $\Lambda$CDM cosmology, assuming a flat Universe, with
$\Omega_m=0.32$, $\Omega_\Lambda=0.68$, and $H_0=68.9\pm 2.4$ km s$^{-1}$
Mpc$^{-1}$.  The reduced $\chi_{\rm dof}^2$ (with 18~degrees of freedom)
for the $R_{\rm h}=ct$ fit is~0.745.  The corresponding value for the
optimal $\Lambda$CDM model (with 16~degrees of freedom) is $\chi_{\rm
dof}^2=0.777$.}}
\end{figure}

It is worth pointing out that the $\Lambda$CDM model optimized for the
cosmic chronometer data alone is quite different from the concordance
model, characterized by the parameter values $H_0=73.8\pm 2.4$ km s$^{-1}$
Mpc$^{-1}$, $\Omega_{\rm m}= 0.27$ and $w_{\rm de}=-1$.  Using the
concordance $\Lambda$CDM parameter values to fit the cosmic chronometer
data yields $\chi^2_{\rm dof}=0.9567$, which is acceptable (since
$\chi^2_{\rm dof}\la1.0)$, but which indicates a noticeably less good fit
than both the fits shown in figure~1.  By comparing the $R_h=ct$ Universe
against the optimized $\Lambda$CDM, rather than against the concordance
model, we are once again presenting $\Lambda$CDM in its best possible
light.

With $n=19$ data points and $k=1$ parameter, the AIC for the optimized
$R_{\rm h}=ct$ Universe is ${\rm AIC}_1=15.41$.  For the optimized
$\Lambda$CDM, with $k=3$, the corresponding value is ${\rm AIC}_2=18.432$.
The magnitude of the difference $\Delta=\allowbreak {\rm AIC}_2-\nobreak
{\rm AIC}_1$, namely $\Delta=3.022$, indicates that $\mathcal{M}_1$~is to
be preferred over~$\mathcal{M}_2$.  According to Equation~(6), the
likelihood of $R_{\rm h}=ct$ (i.e., $\mathcal{M}_1$) being the correct
choice is ${\cal L}(\mathcal{M}_1)=82$ per cent.  For $\Lambda$CDM (i.e.,
$\mathcal{M}_2$), the corresponding value is ${\cal
  L}(\mathcal{M}_2)=18$ per cent.

If one uses the KIC and BIC statistics instead of the AIC, but continues to
weight the models as in Equation~(6), the difference is greater, since $k$
is multiplied by $3$ in the former, and by $\ln n\approx 2.9$ in the
latter.  One finds that ${\rm KIC}_1=16.41$ and ${\rm KIC}_2=21.434$,
yielding ${\cal L}(\mathcal{M}_1)=92.4$ per cent and ${\cal
  L}(\mathcal{M}_2)=7.6$ per cent for $R_{\rm h}=ct$ and $\Lambda$CDM,
respectively.  And for BIC, the results are ${\rm BIC}_1=16.35$ and ${\rm
  BIC}_2=21.27$, yielding ${\cal L}(\mathcal{M}_1)=91.2$ per cent and ${\cal
  L}(\mathcal{M}_2)=8.8$ per cent.

According to all three statistics, the predictions of $R_{\rm h}=ct$ are
more likely than those of $\Lambda$CDM to be closer to the correct
cosmology.  This is notably the case for BIC, for which there is an
accepted interpretation of the magnitude of $\Delta = {\rm BIC}_2 -\nobreak
{\rm BIC}_1$ in~terms of the strength of the evidence against model~2 (Kass
\& Raftery 1995; Tan \& Biswas 2012).  If, as here, $\Delta=4.92$, the
evidence against model~2 (i.e., $\Lambda$CDM) would be judged `positive'
(the positive range for~$\Delta$ extends from 2 to~6, at~which point the
`strong' range begins).

\section{Discussion}
Though a measurement of the cosmic expansion rate using early type galaxies
is subject to several possible systematic errors, the fact that the
inferred values of $H(z)$ are model-independent makes this a highly
desirable and meaningful approach for testing different cosmological
models.  In this paper, we have compared the fits to a data sample drawn
from several sources, and have demonstrated that the $R_{\rm h}=ct$
Universe is more likely than $\Lambda$CDM to account for the observed $H$
versus~$z$ profile.  In~addition, the inferred value of the Hubble constant
$H_0$ is consistent with the rate ($69\pm4$ km s$^{-1}$ Mpc$^{-1}$)
emerging from a fit to the high-$z$ quasar Hubble Diagram (Melia 2012b).

This is rather impressive, given that the former corresponds to a probe of
the local Universe (at $z\la 2$), whereas the latter concerns the cosmic
expansion at high redshift ($z\ga 6$).  In addition, one should not
underestimate the fact that in $R_{\rm h}=ct$, there is only one free
parameter.  By comparison, the standard model of cosmology, with as many as
six, depending on how one parametrizes the dark-energy equation of state,
fails to account for the appearance of high-$z$ quasars at redshift $\ga 6$
(Melia 2013).  This type of comparative analysis therefore supports the
suggestion that $R_{\rm h}=ct$ is closer to the correct cosmology than
is~$\Lambda$CDM\null.  The growing tension between the predictions of
$\Lambda$CDM and the ever improving cosmological data also suggests that
the current standard model may be a useful approximation, but will probably
not endure in the long run.

Recently, however, some criticism has been leveled at the $R_{\rm h}=ct$
cosmology on the basis of several claimed inconsistencies, some
theoretical, others observational (Bilicki and Seikel 2012).  One of the
observational arguments was based on the same cosmic chronometer data we
have addressed in this paper, from which a different conclusion was arrived
at from the one obtained above.  However, these earlier results are
incorrect: simply, because they were not based on a proper statistical
analysis.  Those conclusions appear to have been based on a qualitative
inspection by eye.  But clearly, the results presented here show that such
an approach does not stand up well to a quantitative assessment based on
comparisons of likelihoods.  And since the cosmic chronometer data favor
the $R_{\rm h}=ct$ cosmology over $\Lambda$CDM when using a simple, direct
statistical comparison, any higher-order metric employed with the $H(z)$
data, particularly those designed to test the parametrization of
$\Lambda$CDM, e.g., the decomposition of density into the three specific
components, matter, radiation, and dark energy, cannot be used to
meaningfully constrain the $R_{\rm h}=ct$ Universe.  On the contrary, as we
have shown here, the cosmic chronometer data -- when interpreted
quantitatively -- suggest that the $R_{\rm h}=ct$ Universe is at least as
good as the standard model.

The second observational argument for the criticism
was based on the analysis of Type~Ia SNe.  However, here too the data 
were used incorrectly to arrive at an invalid result.  The use of Type~Ia
supernova data ignored the fact that these were optimized for a pre-assumed
$\Lambda$CDM cosmology.  Therefore, these cannot be used for a comparative
test between different expansion scenarios.  A complete discussion of this
problem has already been published in Melia (2012a), so we shall not
reproduce it here.

The danger of using data optimized for $\Lambda$CDM to test other
cosmologies has also been highlighted recently by an examination of the
Gamma Ray Burst (GRB) Hubble Diagram (Wei et~al.\ 2013).  In this work, the
data were recalibrated separately for each model and, though the results
are quite similar, a comparison of the reduced $\chi^2_{\rm dof}$'s for
$R_{\rm h}=ct$ and $\Lambda$CDM shows that the data clearly favor the
former over the latter.  This result would not have been evident without a
recalibration of the data using the $R_{\rm h}=ct$ expansion history.
Given the preponderance of evidence, it seems likely that when the Type~Ia
SN data are calibrated correctly for each cosmology, $R_{\rm h}=ct$ will
emerge as the more likely of the two to be correct.

Finally, it is worth pointing out that the `theoretical difficulties'
invoked to argue against $R_{\rm h}=ct$ are
based on a failure to comprehend fully Birkhoff's theorem and its
corollary, and the consequence of Weyl's postulate on
Friedmann--Robertson--Walker (FRW) metrics.  It is not the purpose of this
paper to correct this misunderstanding, but since it appears to be an
issue relevant to the interpretation of cosmic chronometer data, we shall
address it here as well.

Birkhoff's theorem and its corollary (Birkhoff 1923) place no limit on
scale, so Bilicki and Seikel's (2012) assertion that the definition of a
Schwarzschild (i.e., a gravitational) radius makes no sense on cosmic
dimensions is without foundation.  Moreover, one does not `define' a
Schwarz\-schild radius, as was claimed; this scale emerges automatically
when one re-writes the metric in terms of observer-dependent coordinates
versus the more commonly used co-moving coordinates (see, e.g., Melia \&
Abdelqader 2009).  Many are perhaps not aware of the fact
that exactly the same phenomenon occurs when writing the spacetime metric
for compact objects.  The distinction arises between a free-falling
observer and the observer at a fixed radius (and hence accelerated)
relative to the central mass.  The former is not aware of the gravitational
radius that emerges only when the metric is written using rulers and clocks
fixed to the latter.  In the cosmological context, we are free-falling
observers when we write the FRW metric using co-moving coordinates.
However, a gravitational radius emerges when we re-write this metric in
terms of an observer's fixed coordinates.

The irony, of course, is that the gravitational radius in cosmology
actually first appeared as far back as 1917, though its meaning was not
then fully appreciated.  de Sitter's (1917) paper on his now famous metric
was originally written in terms of the observer's fixed coordinates, which
included the gravitational radius, since the co-moving coordinates would be
introduced by Friedmann only several years later.  The argument against the
validity of a gravitational horizon in cosmology would therefore imply that
de Sitter space is meaningless on large scales.  This is simply not true.

And since the meaning and validity of the gravitational radius in cosmology
(which, by the way, coincides with the better known Hubble radius) were not
appreciated, the consequence of Weyl's postulate on its permitted rate of
expansion was ignored.  Since the Hubble radius is a `proper'
radius, it has no choice but to expand at a constant rate, as demonstrated
by Melia \& Shevchuk (2012) and (in the more pedagogical treatment) by
Melia (2012c).

\section{Final Remarks}
{\it Euclid} (Laureijs et~al.\ 2011) and {\it BOSS} (Eisenstein
et~al.\ 2011) should provide thousands of passive galaxies at $z>0.5$,
which will significantly improve the accuracy of $H(z)$ at these higher
redshifts.  In concert with this improved statistic, it will
be essential to understand better if the systematic effects, e.g., the
error due to the metallicity and star formation uncertainties, may be
controlled and minimized.  It is crucial to carry out this arduous work,
because these cosmic chronometers are among the few sources that provide us
with model-independent data.  And as we have seen, only such
model-independent data can truly distinguish between competing cosmologies.

We end with a word of caution.  It should be evident from the contents of
this paper how important it is to use only model-independent data in any
comparative analysis between competing cosmologies.  In some cases, it is
simply not possible to avoid the `circularity problem,' in which a model
must be pre-assumed in order to extract the data.  This is certainly the
case in the Type~Ia SN work, but also when dealing with any observations
requiring the use of integrated quantities, such as the luminosity
distance.

An entirely different approach sometimes used to determine $H(z)$ is based
on the identification of Baryon Acoustic Oscillations (BAO) and the
Alcock--Paczynski distortion from galaxy clustering.  That is, instead of
using information on how cosmic time changes with~$z$ (as~is the case for
the data we have used here), this alternative approach measures how
`standard rulers' evolve with redshift.  The results of these two
different methods are sometimes combined to produce an overall $H$
versus~$z$ diagram, but there is a good reason to be wary of this
procedure: whereas the cosmic chronometers produce model-independent data,
the second approach must necessarily assume a particular cosmology and is
therefore model-dependent (Blake et~al.\ 2012).

With the BAO method, the cosmic expansion is measured from the growth of
structure as a function of redshift.  Redshift-space distortions arise
because the recession velocities of galaxies, from which distances are
inferred, include contributions from both the Hubble flow and from the
peculiar velocities driven by the clustering of matter (see, e.g., Hamilton
1998 for a review).  The oscillations are modeled via the non-linear
evolution of both the matter density and velocity fields, which are quite
different between, say, $\Lambda$CDM and $R_{\rm h}=ct$
(Melia \& Shevchuk 2012).  In addition, to compute redshift space
separations for each pair of galaxies given their angular coordinates and
redshifts, one must adopt a cosmological model for the expansion to relate
these quantities to each other.

Unfortunately, this gives rise to a situation not unlike that currently
existing with Type~Ia SNe (Melia 2012a), in which one must simultaneously
optimize at least four nuisance parameters incorporated into the
description of the measurements, along with the free parameters of the
assumed model.  One must therefore avoid the use of such model-dependent
data in any attempts to directly compare fits using $\Lambda$CDM with those
of other models, such as $R_{\rm h}=ct$.  Only the cosmic chronometer data
are truly model-independent and therefore appropriate for this purpose.

\section*{Acknowledgments}
We are grateful to the many workers who spent an extraordinary amount of
effort and time accumulating the data summarized in figure~1.  FM~is also
grateful to Amherst College for its support through a John Woodruff Simpson
Lectureship.  Part of this work was carried out at Purple Mountain
Observatory, Nanjing China.

\end{document}